\documentclass[aps,prl,twocolumn,groupedaddress]{revtex4}
\usepackage{graphicx}
\begin{document}
\title{Anomalous Hall Effect of Calcium-doped Lanthanum Cobaltite Films}
\author{S. A. Baily}
\email[]{baily@uiuc.edu}
\author{M. B. Salamon}
\affiliation{University of Illinois, Department of Physics, Urbana, Illinois}
\date{August 26, 2002}
\begin{abstract}
The Hall resistivity, magnetoresistance, and magnetization of
La$_{1-x}$Ca$_{x}$CoO$_{3}$ epitaxial films with 0.25$\geq$x$\geq$0.4 grown
on lanthanum aluminate were measured in fields up to 7~T.
The x=1/3 film, shows a reentrant metal insulator transition. Below 100~K, the
x=1/3 and 0.4 films have significant coercivity which increases with
decreasing temperature. At low temperature the Hall resistivity remains large
and essentially field independent in  these films, except
for a sign change at the coercive field that is more abrupt than the switching
of the magnetization. A unique magnetoresistance behavior accompanies this
effect. These results are discussed in terms of a percolation picture and the
mixed spin state model for this system. We propose that the low-temperature
Hall effect is caused by spin-polarized carriers scattering off of orbital
disorder in the spin-ordered clusters.
\end{abstract}
\maketitle

\section{Introduction}
In doped lanthanum cobaltites, cobalt ions exist in many different charge
and spin states. The relative number of ions in each state changes
with temperature, giving rise to a complex arrangement of spins and to the
unusual magnetic and transport properties of these compounds. Calcium doped
lanthanum cobaltite has the largest reported anomalous (proportional to
magnetization) Hall effect \cite{AVSamoilov98}. As usual, the maximum
effect occurs near the ferromagnetic transition temperature, where there is
considerable spin disorder. However we report here that the Hall resistivity
remains unusually large at low temperatures. We propose that the
low-temperature Hall effect is caused by spin-polarized carriers scattering
off orbital disorder in the spin-ordered clusters. Further, a unique
magnetoresistance behavior accompanies this effect.

\subsection{Growth, sample quality, etc.}

La$_{1-x}$Ca$_{x}$CoO$_{3}$ films were grown by laser ablation from targets
with x=0.25, 0.33, and 0.4. Polycrystalline targets of each composition were
prepared by pressing and sintering powders produced by a polymeric steric
entrapment method \cite{MHNguyen99}. The films were grown in 150~mTorr of
O$_2$ at 1050~K, and then cooled at 5~K/min to room temperature in 1~atm of
O$_2$. A section of each film was patterned via photolithography and
ion-milled into a five contact Hall pattern. Gold pads were then deposited
for electrical contact. Rutherford backscattering (RBS) was used to
determine the thickness of each film. The thicknesses were 560~\AA, 590~\AA,
and 400~\AA\ for the films with calcium doping of 0.4, 1/3, and 1/4,
respectively. X-ray diffraction indicated epitaxial growth. The growth and
patterning conditions were optimized for the 1/3 calcium doped sample, and
then repeated for the other samples. As part of the optimization process,
the transition temperature was measured by ac magnetic susceptibility,
surface roughness was measured by AFM, and composition was measured by x-ray
photospectroscopy and energy dispersive x-ray analysis. This information was
used along with electrical resistivity and x-ray diffraction data to improve
the laser power/fluence, growth temperature, oxygen pressure, and cooling
rate. The compositions listed for each film are the nominal doping levels.
The RBS data are consistent with these values, but indicate a ten percent
cobalt deficiency for the x=0.4 sample.

\subsection{Measurements}

\begin{figure}[b]
\includegraphics[width=8cm,clip]{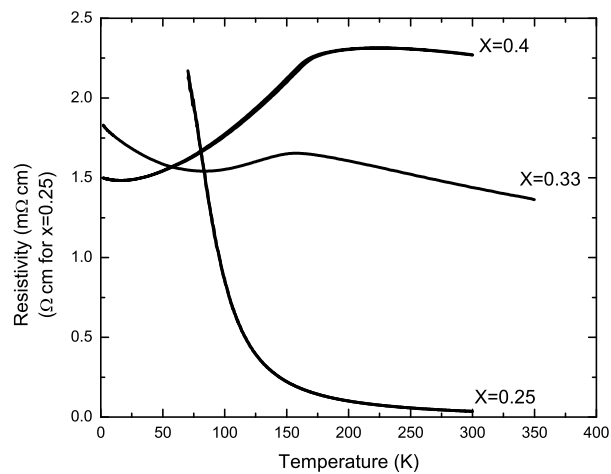}
\caption{La$_{1-x}$Ca$_{x}$CoO$_3$ resistivity vs.\ temperature. The x=1/4 resistivity is scaled by a factor of 1000.\label{RvsT}}
\end{figure}
Magnetoresistivity and Hall resistivity were measured in out-of-plane fields
up to 7~T using a Quantum Design Physical Property Measurement System.
Magnetization measurements were made with the field in the same orientation
with respect to the sample (i.e., perpendicular to the (001) plane) in a
Quantum Design 7~T Magnetic Property Measurement System, so that
demagnetization corrections are not needed to calculate the Hall
coefficient.  A linear temperature independent term is subtracted from the
magnetization vs.\ field data to account for the diamagnetic substrate
signal. The Hall resistivity and magnetoresistance were measured every 0.5~T
from 7~T to -7~T and back. Hall and magnetoresistance measurements are
performed by applying a 37~Hz alternating current to the current leads, and
recording the voltage between all possible combinations of the three voltage
contacts. The weighted sum of these voltages is calculated to yield the
resistivity.  Subtracting two of these voltages yields the Hall voltage plus
a term proportional to the magnetoresistance and the imbalance of the
contacts. Sweeping the field in both directions allows one to average out
this additional term, leaving only the Hall voltage.

\section{Results}

The resistivity for each film is shown in Fig.~\ref{RvsT}. The resistivity
of the x=0.4 sample shows only a slight temperature dependence near room
temperature.  The resistivity drops rapidly below the ferromagnetic
transition temperature (near 180~K). The resistivity of the x=1/3 sample
rises more rapidly at room temperature, then decreases below the
ferromagnetic transition temperature, and increases again below 80~K. The
x=1/4 sample has a much higher resistivity, and is semiconducting at all
temperatures. The behavior of the higher-doped films conflicts with
previously reported data for calcium doped cobaltite films
\cite{AVSamoilov98}, but is similar to that of strontium doped cobaltite
(except that the phase boundaries are at slightly higher doping levels)
\cite{JBGoodenough95}.
\begin{figure}[t]
\includegraphics[width=8cm,clip]{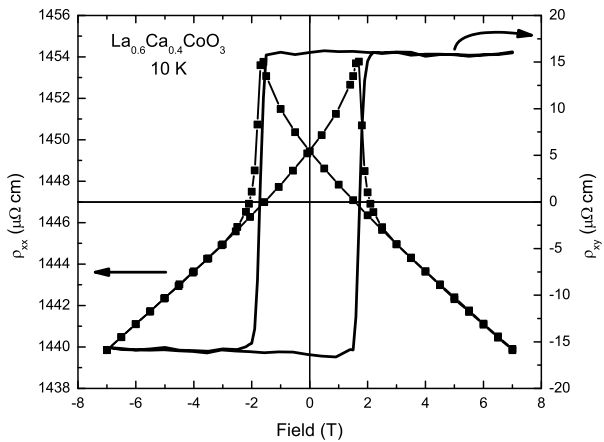}
\caption{La$_{0.6}$Ca$_{0.4}$CoO$_3$ $\rho_{xx}$ and $\rho_{xy}$ vs.\ applied field at 10~K. Only a fixed resistance contribution has been subtracted from $\rho_{xy}$; magnetoresistance contributes a systematic error of up to 3\%.\label{10Kraw}}
\end{figure}

\begin{figure}[b]%
\includegraphics[width=8cm,clip]{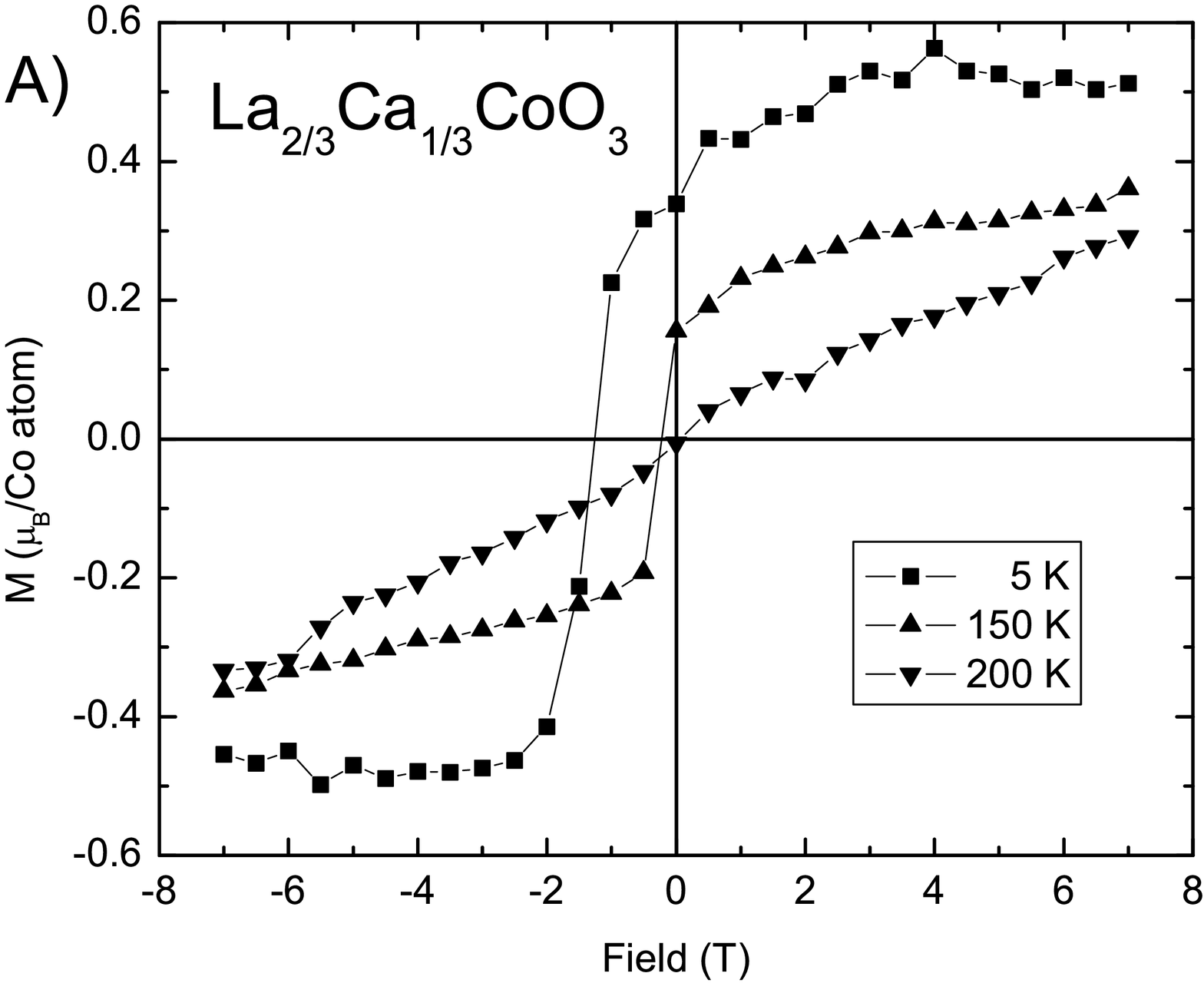}\label{MvsH}\\
\includegraphics[width=8cm,clip]{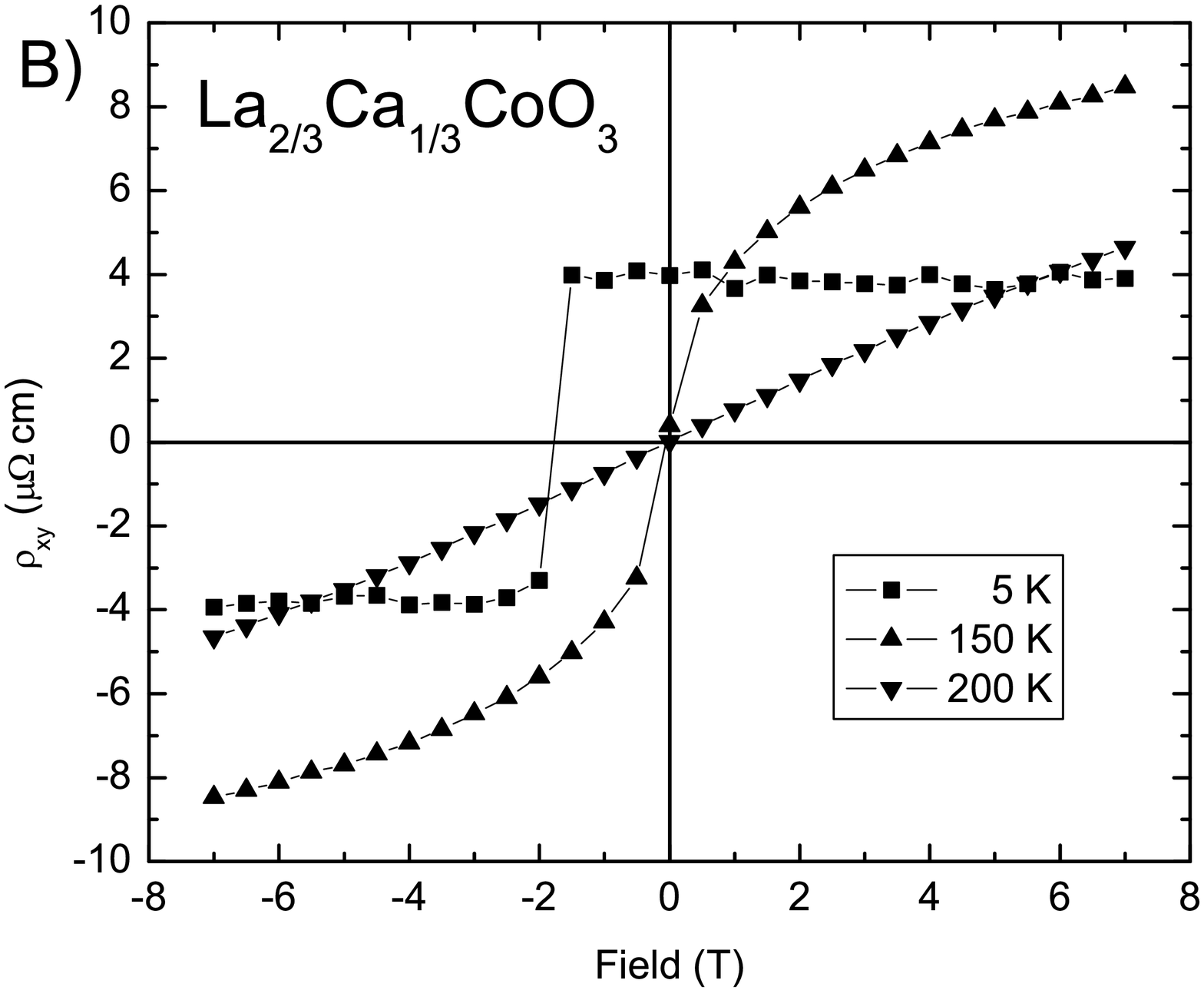}\label{RXYvsH}
\caption{La$_{2/3}$Ca$_{1/3}$CoO$_3$ magnetization and Hall resistivity vs.\ applied field while sweeping from 7~T to -7~T. Both sweep directions are averaged to appropriately remove any magnetoresistance contribution and reduce the noise.\label{MandRXYvsH}}
\end{figure}
Below 100~K, the x=1/3 and 0.4 doped films have
significant coercivity which increases with decreasing
temperature. The resistivity of the x=0.4 
film increases linearly as the applied magnetic field decreases in
magnitude. When the sign of the applied field is changed, the resistivity
continues to increase until switching abruptly at the coercive field. The
magnetoresistance curves are reversible at fields larger than the coercive
field. This magnetoresistance is reminiscent of that of magnetic
multilayers, except that the shape of the curves is different and cobaltite
shows this behavior in much higher fields. Fig.~\ref{10Kraw} shows a
striking example of this behavior.  The Hall data shown in Fig.~\ref{10Kraw}
includes both sweep directions with the average value subtracted. Thus
removing the imbalance of the contacts, but neglecting a contribution of up
to 3\% from the magnetoresistance. The relative magnitude of the
magnetoresistance of the 0.4 calcium doped film is temperature independant.
The x=1/3 film behaves in a similar manner except that the magnitude of the
magnetoresistance changes with temperature, and has opposite sign below
60~K. The anomalous Hall resistivity of the higher doped samples peaks near
the ferromagnetic transition temperature, but still remains quite
significant at low temperatures. Although the Hall resistivity switches at
the coercive field, the change is much more abrupt than that of the
magnetization (Fig.~\ref{MandRXYvsH}). The Hall resistivity is nearly field
independent about and below the switching field, depending on the sign of
the magnetization, but not its magnitude.

\section{Discussion}

Calcium doped cobaltite behaves similar to strontium doped cobaltite, except
that the phase boundaries are at slightly higher doping levels. In M. A.
Se\~{n}ar\'{\i}s-Rodr\'{\i}guez and J. B. Goodenough's model for strontium
doped cobaltite, cobalt atoms exist in two different ionization states, and
many different spin states \cite{JBGoodenough95}. The intermediate-spin
trivalent and low-spin tetravalent cobalt ions have 5 electrons in the
t$_{2g}$ levels. Jahn-Tellar distortion splits the t$_{2g}$ levels into
doublet and singlet states. The doublet t$_{2g}$ state is degenerate because
it contains 3 electrons, and this results in large spin-orbit
coupling \cite{MRIbarra98}\cite{RGanguly99}. At higher calcium doping levels,
double-exchange hopping between aligned sites creates a conducting
percolation path through an insulating matrix. This path is ferromagnetic
with large coercivity below 100~K. The insulating matrix has a magnetization
which simply follows the applied magnetic field (i.e., paramagnetic
behavior) \cite{SABaily2002}. For the 0.4 calcium doped sample, the
effective width of the percolation path is proportional to the degree of
alignment of the spin of the matrix with that of the percolation path. When
the field is high, the path widens, because the spins tend to line up, and
the path is narrowest when the matrix has spins antiparallel to those of the
percolation path. The electrons on the percolating cluster are fully
polarized, and therefore immune to changes in field (once the percolation
cluster itself is saturated). There is spin order on the cluster, but
significant orbital disorder \cite{JBGoodenough95}\cite{MRIbarra98}%
\cite{RGanguly99}. The Hall effect, then, results from spin-orbit scattering
of the spin-polarized electrons off the orbital disorder \cite{PNozieres73}.
The sign of the Hall effect depends only on the magnetization direction of the
half-metallic percolating pathway. Below 80~K the 1/3 calcium doped sample has
ferromagnetic order, but does not percolate electrically. Thermally
populated spin states apparently contribute to the percolation path. As the
temperature decreases the number of low-spin cobalt states in the insulating
matrix increases, suppressing the conductivity at low temperature. It is
possible that decreasing the number of thermally populated spin states
creates regions where minority spins are scattered more strongly than
majority spins and other regions where the opposite is true. These
differential spin scattering rates may change as a function of temperature.
If this is the case then it is reasonable for the shape of the
magnetoresistance curves to remain the same while changing in sign and
magnitude. Similar behavior has been observed in multilayer films when the
differential spin scattering rate differs in alternating magnetic layers
\cite{JMGeorge94}. At higher doping the percolation path remains stable and
as a result there is little change in the magnetoresistance ratio.

\begin{acknowledgments}
The authors would like to thank B. Rosczyk and W. M. Kriven for providing
the powder for targets, and D. J. Van Harlingen and W. K. Neils for the use
of their laser ablation chamber. This
material is based upon work supported by the U.S. Department of Energy,
Division of Materials Sciences under Award No. DEFG02-91ER45439, through the
Frederick Seitz Materials Research Laboratory at the University of Illinois
at Urbana-Champaign.
\end{acknowledgments}

\bibliography{mmm02}

\begin{thebibliography}{8}
\expandafter\ifx\csname natexlab\endcsname\relax\def\natexlab#1{#1}\fi
\expandafter\ifx\csname bibnamefont\endcsname\relax
  \def\bibnamefont#1{#1}\fi
\expandafter\ifx\csname bibfnamefont\endcsname\relax
  \def\bibfnamefont#1{#1}\fi
\expandafter\ifx\csname citenamefont\endcsname\relax
  \def\citenamefont#1{#1}\fi
\expandafter\ifx\csname url\endcsname\relax
  \def\url#1{\texttt{#1}}\fi
\expandafter\ifx\csname urlprefix\endcsname\relax\def\urlprefix{URL }\fi
\providecommand{\bibinfo}[2]{#2}
\providecommand{\eprint}[2][]{\url{#2}}

\bibitem[{\citenamefont{Samoilov et~al.}(1998)\citenamefont{Samoilov, Beach,
  Fu, Yeh, and Vasquez}}]{AVSamoilov98}
\bibinfo{author}{\bibfnamefont{A.~V.} \bibnamefont{Samoilov}},
  \bibinfo{author}{\bibfnamefont{G.}~\bibnamefont{Beach}},
  \bibinfo{author}{\bibfnamefont{C.~C.} \bibnamefont{Fu}},
  \bibinfo{author}{\bibfnamefont{N.-C.} \bibnamefont{Yeh}}, \bibnamefont{and}
  \bibinfo{author}{\bibfnamefont{R.~P.} \bibnamefont{Vasquez}},
  \bibinfo{journal}{Phys.\ Rev.\ B} \textbf{\bibinfo{volume}{57}},
  \bibinfo{pages}{R14032} (\bibinfo{year}{1998}).

\bibitem[{\citenamefont{Nguyen et~al.}(1999)\citenamefont{Nguyen, Lee, and
  Kriven}}]{MHNguyen99}
\bibinfo{author}{\bibfnamefont{M.~H.} \bibnamefont{Nguyen}},
  \bibinfo{author}{\bibfnamefont{S.~J.} \bibnamefont{Lee}}, \bibnamefont{and}
  \bibinfo{author}{\bibfnamefont{W.~M.} \bibnamefont{Kriven}},
  \bibinfo{journal}{J.\ Mater.\ Res.} \textbf{\bibinfo{volume}{14}},
  \bibinfo{pages}{3417} (\bibinfo{year}{1999}).

\bibitem[{\citenamefont{Se{\~n}ar{\'{\i}}s-Rodr{\'{\i}}guez and
  Goodenough}(1995)}]{JBGoodenough95}
\bibinfo{author}{\bibfnamefont{M.~A.}
  \bibnamefont{Se{\~n}ar{\'{\i}}s-Rodr{\'{\i}}guez}} \bibnamefont{and}
  \bibinfo{author}{\bibfnamefont{J.~B.} \bibnamefont{Goodenough}},
  \bibinfo{journal}{J. Solid State Chem.} \textbf{\bibinfo{volume}{118}},
  \bibinfo{pages}{323} (\bibinfo{year}{1995}).

\bibitem[{\citenamefont{Ibarra et~al.}(1998)\citenamefont{Ibarra, Mahendiran,
  Marquina, Garc{\'{\i}}a-Landa, and Blasco}}]{MRIbarra98}
\bibinfo{author}{\bibfnamefont{M.~R.} \bibnamefont{Ibarra}},
  \bibinfo{author}{\bibfnamefont{R.}~\bibnamefont{Mahendiran}},
  \bibinfo{author}{\bibfnamefont{C.}~\bibnamefont{Marquina}},
  \bibinfo{author}{\bibfnamefont{B.}~\bibnamefont{Garc{\'{\i}}a-Landa}},
  \bibnamefont{and} \bibinfo{author}{\bibfnamefont{J.}~\bibnamefont{Blasco}},
  \bibinfo{journal}{Phys.\ Rev.\ B} \textbf{\bibinfo{volume}{57}},
  \bibinfo{pages}{R3217} (\bibinfo{year}{1998}).

\bibitem[{\citenamefont{Ganguly et~al.}(1999)\citenamefont{Ganguly,
  Gopalakrishnan, and Yakhmi}}]{RGanguly99}
\bibinfo{author}{\bibfnamefont{R.}~\bibnamefont{Ganguly}},
  \bibinfo{author}{\bibfnamefont{I.~K.} \bibnamefont{Gopalakrishnan}},
  \bibnamefont{and} \bibinfo{author}{\bibfnamefont{J.~V.}
  \bibnamefont{Yakhmi}}, \bibinfo{journal}{Physica B}
  \textbf{\bibinfo{volume}{271}}, \bibinfo{pages}{116} (\bibinfo{year}{1999}).

\bibitem[{\citenamefont{Baily et~al.}(2002)\citenamefont{Baily, Salamon,
  Kobayashi, and Asai}}]{SABaily2002}
\bibinfo{author}{\bibfnamefont{S.~A.} \bibnamefont{Baily}},
  \bibinfo{author}{\bibfnamefont{M.~B.} \bibnamefont{Salamon}},
  \bibinfo{author}{\bibfnamefont{Y.}~\bibnamefont{Kobayashi}},
  \bibnamefont{and} \bibinfo{author}{\bibfnamefont{K.}~\bibnamefont{Asai}},
  \bibinfo{journal}{Appl.\ Phys.\ Lett.} \textbf{\bibinfo{volume}{80}},
  \bibinfo{pages}{3128} (\bibinfo{year}{2002}).

\bibitem[{\citenamefont{Nozi{\`e}res and Lewiner}(1973)}]{PNozieres73}
\bibinfo{author}{\bibfnamefont{P.}~\bibnamefont{Nozi{\`e}res}}
  \bibnamefont{and} \bibinfo{author}{\bibfnamefont{C.}~\bibnamefont{Lewiner}},
  \bibinfo{journal}{J. Phys.} \textbf{\bibinfo{volume}{34}},
  \bibinfo{pages}{901} (\bibinfo{year}{1973}).

\bibitem[{\citenamefont{George et~al.}(1994)\citenamefont{George, Pereira,
  Barth{\'e}l{\'e}my, Petroff, Steren, Duvail, Fert, Loloee, Holody, and
  Schroeder}}]{JMGeorge94}
\bibinfo{author}{\bibfnamefont{J.~M.} \bibnamefont{George}},
  \bibinfo{author}{\bibfnamefont{L.~G.} \bibnamefont{Pereira}},
  \bibinfo{author}{\bibfnamefont{A.}~\bibnamefont{Barth{\'e}l{\'e}my}},
  \bibinfo{author}{\bibfnamefont{F.}~\bibnamefont{Petroff}},
  \bibinfo{author}{\bibfnamefont{L.}~\bibnamefont{Steren}},
  \bibinfo{author}{\bibfnamefont{J.~L.} \bibnamefont{Duvail}},
  \bibinfo{author}{\bibfnamefont{A.}~\bibnamefont{Fert}},
  \bibinfo{author}{\bibfnamefont{R.}~\bibnamefont{Loloee}},
  \bibinfo{author}{\bibfnamefont{P.}~\bibnamefont{Holody}}, \bibnamefont{and}
  \bibinfo{author}{\bibfnamefont{P.~A.} \bibnamefont{Schroeder}},
  \bibinfo{journal}{Phy.\ Rev.\ Lett.} \textbf{\bibinfo{volume}{72}},
  \bibinfo{pages}{408} (\bibinfo{year}{1994}).

\end{thebibliography}
\end{document}